# Massively Parallel Amplitude-Only Fourier Neural Network

*Mario Miscuglio[1], Zibo Hu[1], Shurui Li[2], Jonathan George[1], Roberto Capanna[3], Philippe M. Bardet[3], Puneet Gupta[2], and Volker J. Sorger[1].\**

\* *sorger@gwu.edu*

[1]*Deptartment of Electrical and Computer Engineering, George Washington University, Washington DC, DC, USA*
[2]*Deptartment of Electrical and Computer Engineering, University of California, Los Angeles, CA, USA*
[3]*Department of Mechanical and Aerospace Engineering, George Washington University, Washington, DC, USA*

**Abstract –**
**Machine-intelligence has become a driving factor in modern society. However, its demand outpaces the underlying electronic technology due to limitations given by fundamental physics such as capacitive charging of wires, but also by system architecture of storing and handling data, both driving recent trends towards processor heterogeneity. Task-specific accelerators based on free-space optics bear fundamental homomorphism for massively parallel and real-time information processing given the wave-nature of light. However, initial results are frustrated by data handling challenges and slow optical programmability. Here we introduce a novel amplitude-only Fourier-optical processor paradigm capable of processing large-scale ~(1,000 × 1,000) matrices in a single time-step and 100 microsecond-short latency. Conceptually, the information-flow direction is orthogonal to the two-dimensional programmable-network, which leverages $10^6$-parallel channels of display technology, and enables a prototype demonstration performing convolutions as pixel-wise multiplications in the Fourier domain reaching peta operations per second throughputs. The required real-to-Fourier domain transformations are performed passively by optical lenses at zero-static power. We exemplary realize a convolutional neural network (CNN) performing classification tasks on 2-Megapixel large matrices at 10 kHz rates, which latency-outperforms current GPU and phase-based display technology by one and two orders of magnitude, respectively. Training this optical convolutional layer on image classification tasks and utilizing it in a hybrid optical-electronic CNN, shows classification accuracy of 98% (MNIST) and 54% (CIFAR-10). Interestingly, the amplitude-only CNN is inherently robust against coherence noise in contrast to phase-based paradigms and features an over 2 orders of magnitude lower delay than liquid crystal-based systems. Beyond contributing to novel accelerator technology, scientifically this amplitude-only massively-parallel optical compute-paradigm can be far-reaching as it de-validates the assumption that phase-information outweighs amplitude in optical processors for machine-intelligence, such as for information processing at the network-edge, in data centres, or for pre-processing information or filtering towards near real-time decision making.**

## Introduction

In the recent years, deep learning has thrived due to its ability to learn patterns within data and perform intelligent decisions, superior in some cases to human [1–3]. Convolution neural networks (CNN) lie at the heart of many emerging machine learning applications, especially those related to the analysis of visual imagery. From a neural network (NN) point of view, a CNN extracts specific features of interest, using linear mathematical operations – convolutions – which combine two pieces of information, namely feature map and kernel, to form a third function (transformed feature map). Interestingly, these convolution layers are responsible for consuming the majority (~80%) of the compute resources during inference tasks[4]. In fact, the convolution between a feature map ($n \times n$) and a kernel ($k \times k$) requires a computational complexity of $O(n^2k^2)$ in the real spatial domain, hence without performing any transformation. This results in a significant latency and computational power consumption, especially for datasets comprising appreciably large feature maps, or requiring deep CNNs for achieving high accuracy[5], even when the network has been trained and the memory initialized. For this purpose, data-parallel specialized architectures such as Graphic Processing Units (GPUs) and Tensor Processing Units (TPUs), providing a high-degree of programmability, deliver dramatic performance gains compared to general-propose processors.

When used to implement deep NN performing inference on large two-dimensional data sets such as images, TPUs and GPUs are rather power-hungry and require a long computation time (> tens of ms), which is function of the complexity of the task and accuracy required, which translates into manifold operations with complex kernel and larger feature map.

As it stands, improving computational efficiency of CNNs is still a challenge, due to the widespread relevance to many applications. Therefore, it is necessary to reinvent the way current computing platforms operate, replacing sequential and temporized operations, and related continuous access to memory with massively parallelized yet distributed dynamical units, pushing towards efficient post-CMOS compute paradigms and system implementations. The intrinsic parallelism and simultaneous low-energy consumption make free space optics a particularly attractive candidate for deep-learning, computing, and, particularly for image classification and pattern recognition using CNNs in real-time (low latency). In this context, as late as the 1960s, optical filtering and correlations, relying on spatial Fourier transform of images in frequency domain, were used to extrapolate similarity (specific features) between images and signatures[6]. Subsequently, research groups built optical correlator and convolution processors[7,8], with competitive performance for that period, although the tremendous advancement of digital electronics frustrated these efforts. However, early successes of such optical processors did not step beyond prototypes stages due to the inability to feed these potentially high-throughput (~POPS/s) processors sufficiently with data front-end.

Increased data volume and parallel computation requirements together with recent advances in digital display technology poise new opportunities for massively parallel optical accelerators. Optical free-space systems offer processing large matrices (several Megapixels), and the CNN-required convolutions can be performed as simpler pointwise multiplications in the Fourier domain where domain crossings (from real- to Fourier space, and inverse) are performed passively in a Fourier-optics $4f$ system. However, the high parallelism and inherent operations provided by the nature of optical signal is confronted by the rigidity of the current optical tools which lack high-speed programmability. For instance, recent optical systems, used as convolutional layer performing inference after being trained, rely on fixed kernels, realized as 3D-printer manufactured diffractive masks[9], or slowly varying (10's Hz) Spatial Light Modulators (SLMs)[10]. On the other hand, state-of-the-art high-speed (GHz) programmable metasurfaces and tunable optical phased array are still limited in terms of matrix resolution and phase contrast[11,12].

Here, we introduce and experimentally demonstrate a novel compute paradigm based on amplitude-only (AO) electro-optical convolutions between large matrices or images using kHz-fast reprogrammable high resolution digital micromirror devices (DMDs), based on two stages of Fourier Transforms (FT), without the support of any interferometric scheme. Low-power laser light is actively patterned by electronically configured DMDs in both the object and Fourier plane of a $4f$ system, encoding information only in the amplitude of the wave-front. By individually controlling the 2 million programmable micromirrors, with a resolution depth of 8 bit and a speed of 1,031 Hz (~20 kHz with 1 bit resolution), it is possible to achieve reprogrammable operations for (near) real-time, which is about 100x lower system latency with respect to current GPU accelerators (SLM-based systems[10]) image processing, with a maximum throughput of 4-Peta operations per second at 8 bit resolution, emulating on the same platform multiple convolutional layers of a NN.

Additionally, this study tackles an enduring scientific question of AO image correlators presented by Oppenheim and Lim[13] in 1984, who demonstrated that from an image processing perspective, phase information is considerably more important than the amplitude information in the transmission of a continuous tone picture for preserving its visual intelligibility[14]. Leveraging on the robustness of the NN, achieved through hardware-specific training, we show that it is possible to overcome the loss of information related to phase of the modulated radiation which enables performing intelligent classification in an opportunely trained CNN and concurrently achieving high accuracy (MNIST and CIFAR-10 classification) and throughput (10000 Conv/s of ~2,000 × 1,000 large matrices). This architecture experimentally validates the power of an AO 4*f* system optical computing paradigm and further opens up the NN architectures with components that are readably available for parallelly performing intelligent tasks in near real-time, such as in free-space communication[15] in data centres for processing data locally at the edge of the network, without communicating across long routes to data centres or clouds.

## Results

The system architecture typology for realizing the Amplitude Only Fourier Filter (AO-FF) layer for performing filtering is synergistically realized in optics;[16] a coherent optical image processor is based on a 4*f* system, in which there are four lens focal distances *f* separating the object from the image plane, intercalated by two Fourier transforming lenses (**Fig. 1a**). This setup is composed of an input (object) plane, the processing (Fourier) plane, and the output (image) plane. The to-be-processed data and the kernel, which filters them in the Fourier plane, are spatially modulated according an electro-optic transduction. Conceptually, such a free-space approach enables three-dimensional parallelism, which is elegant, since it decouples in-plane (x, y-direction) programmability (here provided by the DMD), from the direction of the information flow (z-direction).

With the presumption that phase information is more relevant than amplitude information[13], other 4*f* implementations rely on phase modulation based on Spatial Light modulators (SLMs) [10]. SLMs exploit pixelwise phase retardation introduced by the variation of the effective refractive index through orientation of birefringent liquid crystals to which a voltage is applied. On the contrary, for our implementation, this transduction is achieved through a DMD, belonging to the family of Micro-opto-electro-mechanical system (MOEMS). They consist of micromirror arrays which impose a spatially varying light intensity modulation by rapidly tilting individuals micromirrors, which deflect input light. In details, each pixel of a DMD comprises of a tilting mirror and a memory unit storing the pattern to be reproduced; the mirror flips according to the digital value stored in memory to let the light either pass or being deflect. Assuming the same pixel resolution (2MPx or 4K), readily available DMDs are characterized by at least two orders of magnitude faster (tens of kHz) settling speed compared to SLM (tens of Hz), making them a promising platform for optical computing, thus subject of this study.

In our optical engine (**Fig. 1b**), collimated low-power laser light (633 nm, HeNe Laser) is expanded to uniformly interest the entire active area of the first DMD in the object plane, which, by independently tilting each micromirror of its array according to a pre-loaded image, defines the input image (feature map). The DMD in the object plane is oriented with a 22.5° tilting angle with respect to the normal incidence and rotated in-plane by 45° degrees. Light reflected from the DMD is Fourier-transformed passing through the first Fourier lens at one focal length, $f$, apart from the first DMD in the object plane. The pattern in the second DMD, with specular orientation with respect to the first one, acts as a spatial mask in the Fourier plane, opportunely selecting the spatial frequency components of the input image. The frequency filtered image is inverse Fourier transformed into the real space by the second Fourier lens and imaged by a high-speed camera (**Fig. 1b**). Both FT transformation steps are performed entirely passively, i.e. zero-static power consumption, which is in stark contrast to performing convolutions as dot product multiplications in electronics[5].

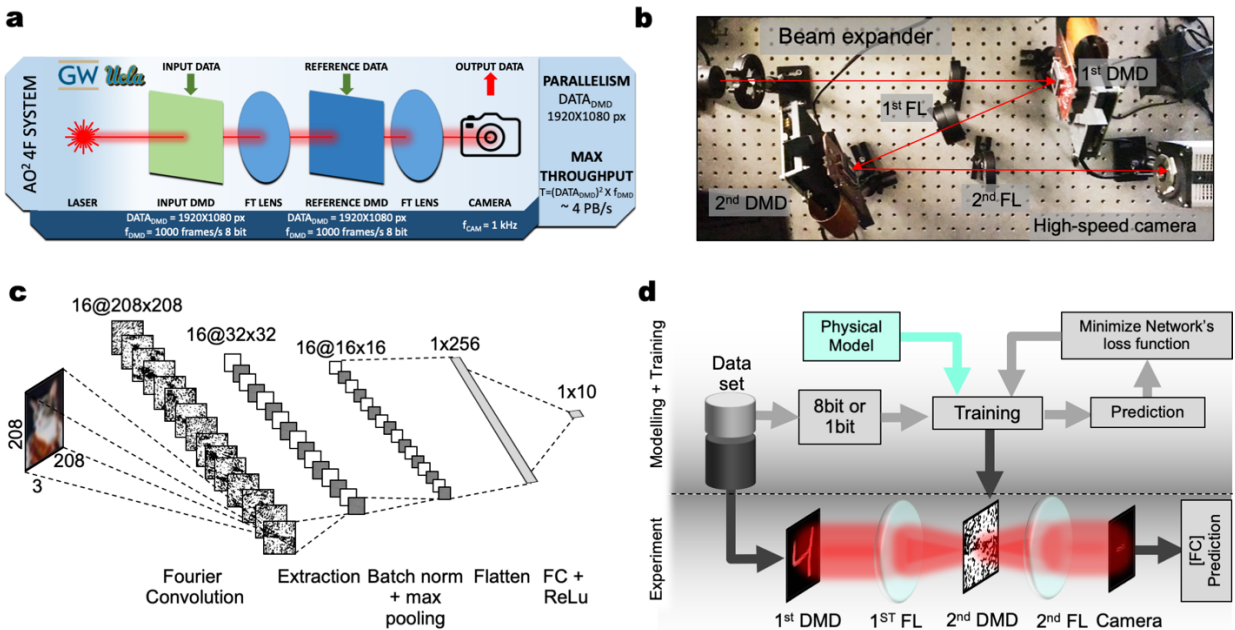

***Figure 1. Amplitude only Fourier Neural Network****. **a** Schematic representation of a 4f system based on a Digital Micromirror Devices (DMDs). The amplitude of a low power light source is modulated according to a pattern (input data). The image so generated is Fourier transformed and multiplied with a reference data in the Fourier plane of a 4f system, affecting only its amplitude. The result of the product is inverse transformed, and the square of its intensity is imaged by the camera showcasing the same spatial resolution (pixel size and pitch) of the DMDs. **b** Experimental implementation of the amplitude only Fourier filter based on a DMD 4F system. **c** Convolutional Neural Network (CNN) structure for CIFAR 10 dataset. The optical Amplitude only Fourier filter is used as convolution layer, with the subsequent layers realized electronically. The kernels obtained during physically meaningful training are loaded in the 2nd DMD. After a convolution layer a nonlinear thresholding is applied to the output (Rectified Linear unit function) and are pooled together. A flatten layer collapses the spatial dimensions of the output into the channel*

*dimension to which follows a fully connected layer and a nonlinear activation function. **d** Flow-chart of the training process. Physical model of the amplitude only Fourier filter layer is used for training the entire CNN. (**c**), obtaining the weights for the kernel to be loaded in the 2ⁿᵈ DMD of the convolution layer. Experimentally obtained results of the Amplitude Only Fourier filtering are fed to the FC layer for performing the final prediction on unseen data.*

On the system level, a computer loads the input image as well as the kernel (1920x1080, 8 bit, 1000 Hz) stored in its memory to the DMDs by means of a HDMI cable or directly generated through an FPGA (Virtex 7), which connects to the Digital Light Processing (DLP) boards (Texas Instrument) of the 2 DMDs through a serial connection, aiming to reduce the latency in providing the signals and allowing for processing while streaming data. Consequently, the AO Fourier filtered images are detected with a charge-coupled-device (CCD) camera (1000 Frames/s with 8bit resolution) connected through PCI-express to a unified system interface which can store the data or process it implementing other NN tasks, such as max pooling, activation function and fully connected layer. For emulating deeper neural networks which comprise multiple layer, the resulting image could be potentially loaded in the 1ˢᵗ DMD (see more details in Section 1 of the supplementary information in Ref. 35).

Considering the abovementioned specifications, the system leverages: (1) the vast parallelism given by the high pixel resolution of the camera and DMDs (2Mpx); (2) inherent and completely passive operations due to the wave-nature of the optical radiation, which allows for passive Fourier transforming exploiting lenses (Fresnel's integral) and pixel-wise multiplication in the Fourier plane (Huygens' Principle); (3) order of magnitude faster update rates compared to SLMs based on liquid crystals; thus (4) enabling a nominal throughput equivalent to 4 Peta operations per second performed by space domain convolution operations (sliding window), with a resolution given by the DMDs (1920x1080 at 8bit), updating at a frequency of ~1 kHz and with a CCD camera acquisition frame rate of 1 kHz. It is worth to stress that, unlike other implementations[9,17] in which the kernels are fixed phase masks (diffractive elements or optical transparencies) and cannot be adjusted after training without physically replacing it, in our convolutional layer both feature maps and kernels can be updated at the same high rate (10 kHz). This can be particularly advantageous for emulating on the same hardware, deeper CNN architecture, which comprises multiple convolutional layers, in which batch normalization and max pooling are performed in the electrical domain. Notice that our convolutional layer already provides a straightforward nonlinearity (threshold) without the need of all optical nonlinearities as proposed by other schemes[18], since after the linear operation computed in the spatial frequency filtering (convolution) performed by the 4$f$ system, at the image plane, the electric field intensity associated to light is squared ($x^2$ function), when detected by the camera.

The proposed AO Fourier filter (AO-FF) could be particularly useful in systems in which the input images are already encoded in a coherent radiation (first DMD is absent); for instance, the AO-FF can detect images within images (such as in steganography and optical illusions as shown in Section 2 Ref. 35), demonstrating an immediate use in augmented visual perception or in classification of complex pattern, such as in iris recognition 8-bit scans.

Interestingly, spatial frequency filtering performed by a DMD is insensitive to the phase information. It is well established that full-field control could be achieved but here it is not desired. In 1963, in fact, Van der Lugt proposed a way to achieve plane-frequency mask which retains effective phase and amplitude control in spite of using just absorption patterns[6], by exploiting Fourier holograms of the input image. Other full-field spatial control can be achieved through several interferometric schemes[19], such as Rayleigh or Mach-Zehnder interferometer, Lee holograms[20], superpixel[21] and more recent high-precision methods[22] and NN-based holographic reconstruction[23]. The full control over the optical field, while being advantageous in terms of image processing, although, comes at a cost of; (1) increased complexity of the system, requiring additional optics and cumbersome alignments; (2) reduction of the total dimension of the phase mask or need to corrective measurements and consequent drop of the overall parallelism. For these reasons, unlike other demonstrations[24], we deliberately decide to train the CNN to account for the information loss related to phase, and for the imprecise reconstruction of the images, while performing convolutions.

The designed CNN architecture consists of a single convolution layer in which sets of kernels are convolved with the input images. The convolutional layers are usually intercalated by pooling layer, which reduces the matrix dimensionality followed by nonlinear thresholding. Typical multilayer-CNNs comprise layers of convolutional nodes followed by layers of fully connected nodes. Here, we use our experimental optical AO Fourier convolutional layer, whose output is pooled together, followed by a fully connected layer and nonlinear thresholding, both performed electronically. The convolutional layer has 16 nodes and each convolutional node uses a 208×208 kernel. The kernel parameters comprising the weights that are learned during the training procedure (**Fig. 1c**). The CNN is trained using PyTorch, which is agnostic to the optics hardware.

Therefore, it uses a set of functions, which exhaustively describe the Fourier convolution layer in order to accurately simulate the physical system. We adopt the concept of FFT-based Fourier domain training[25], together with the refined hardware model to accurately simulate the complete process and learn the kernel weights during training. The kernel values, which are the learnable parameters of the convolutional layer, are initialized directly in the Fourier domain. By doing so, the kernels do not need to be transformed into Fourier domain such as required in [26,27], which matches our physical model well. For fully utilizing the maximum update speed of the DMD we restrict the kernel values to be real and binary, therefore in the training a custom binarization step is needed. The CNN is trained using two classic datasets for image recognition to demonstrate the learning capability of this system as well as benchmarking it, namely MNIST dataset of hand-written digits and CIFAR-10, a more challenging image classification problem. The trained kernel is used as an input pattern in the free space 4*f* system and the results of the convolutions are used for validating the physical model and for eventual successive training of the fully connected (FC) NN (**Fig. 1d**).

For obtaining a correct training and consequently highly accurate inference when performing convolution using the optical hardware, the physical model embedded into the training phase needs to accurately describe the coherent optical engine including its analogue-computing approximations and inaccuracies (More details Section 3 Ref. 35).

In order to validate the model and compare the results with the experimental realization of the optical engine, at first, we filter, by way of example, the 8 bit image of GWU mascot (the Colonial), using different spatial frequency filters[35]. The results of the convolution obtained through the model and the experimental realization highlights a qualitative and quantitative agreement obtaining high (>0.7 for all the kernel except low-pass filter) Structural SIMilarity (SSIM), which is related to the image degradation as perceived change in structural information, and extremely low absolute errors, showcased by <0.1 Root Mean Square Error (RMSE). (More details in Section 4 Ref. 35)

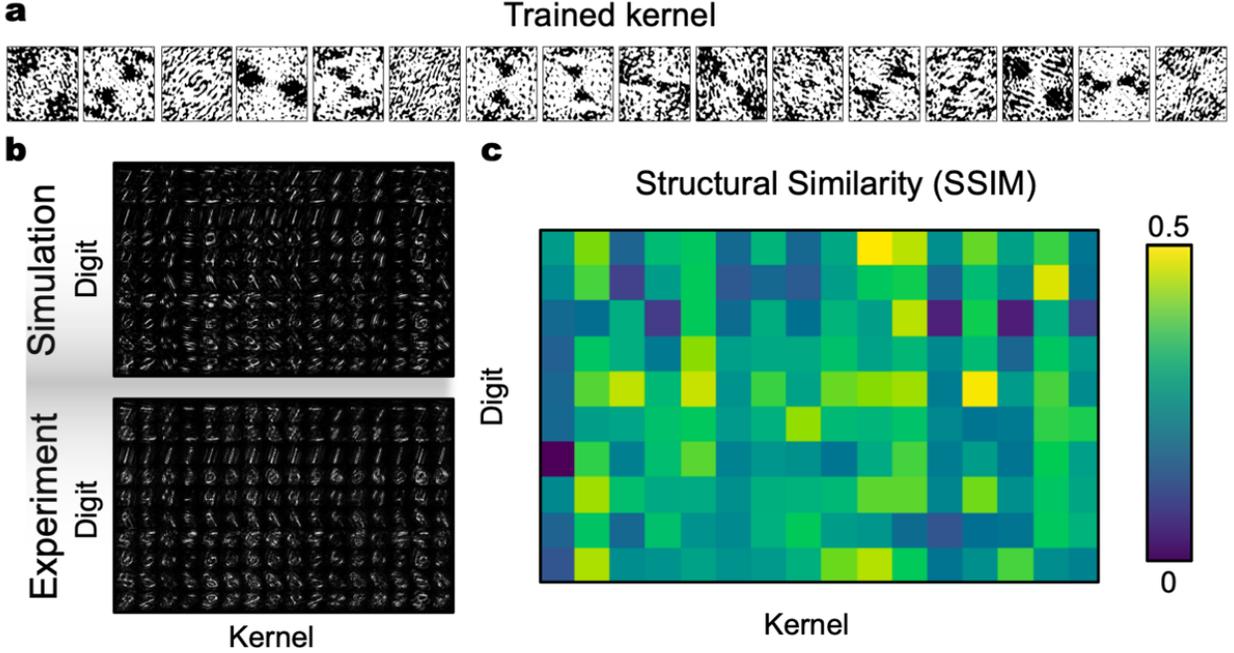

*Figure 2. Experimental testing of MNIST classifier. **a** Kernel obtained during training of the Fourier Neural Network for the classification of handwritten digits (MNIST dataset). **b** Output result of the emulated and experimental implementation of the first layer for different kernels (x-axis) and input images (y-axis). **c** Structural Similarity map which compares the output obtained experimentally and those obtained during emulation for different digits (y-axis) and kernels (x-axis). We used the experimental output for training only the fully connected layer in order to compensate the discrepancies and improving the accuracy of the inference.*

Leveraging on the massive amount of parallelism available in optical hardware (2MPx), the AO Fourier-based convolutional layer can be further parallelized if the input images (208x208 pixel) are smaller compared to the resolution offered by the DMD and the camera. In our experiment, we tiled in the input plane and batch-process up to 46 images using the same kernel in the Fourier plane. Alternatively, the same input can be simultaneously filtered by multiple kernels; in this case the Fourier transformed image is directed to different (non-overlapping) portion of the DMD (or different DMDs) in the Fourier plane using opportune beam splitters, array of mirrors and well-dimensioned lenslet array. Ultimately each product is inverse Fourier transformed (using a second lenslet array) and imaged by different sensors. The filtered images can be integrated by the same sensor, performing dimensionality reduction. For additional information regarding the experimental implementation of the parallelization schemes see Section 5 Ref. 35.

After the model validation and establishment of parallelization schemes, to demonstrate the performance of the AO-FNN, we first trained the processor as an image classifier, performing automated classification of handwritten digits (MNIST). For this task, we train a 1-layer convolutional layer, followed by a FC layer, with 55,000 images (5000 validation images) from the MNIST (Modified National Institute of Standards and Technology) handwritten digit database. The input digits are encoded as amplitude and the network is trained to obtain the kernels (16, 208 × 208 binary images) to multiply in the Fourier plane, to be fed to the second DMD (**Fig. 2a**). More details on the training are provided in Section 6 of Ref. 35.

After the training, the network was blind tested, adopting the obtained kernel, using unseen images from the MNIST test dataset (not used as part of the training/validation), achieving 98% classification accuracy (**Table 1**). At this stage, for validating the hardware implementation, we perform convolutions between the kernels and unseen feature maps using the optical engine. The results of the emulated and experimental convolution layer are compared in terms of transformed feature maps and classification accuracy. Since our simulation model already considers some nonidealities of the optical hardware, the convolution results of the hardware implementation match the simulation result quite well qualitatively, their shapes are almost identical (**Fig. 2b**). Although, the match is not perfect quantitatively, highlighted by a lower SSIM (**Fig. 2c**). This is due to several concurring factors including a) small misalignment in the optical setup, b) model which takes into account unphysical reflection of grid boundaries, c) non-ideal camera dynamic range. The exact pixel values of hardware results differ from the simulation results, thus if the convolution results obtained using the optical hardware are fed into a fully connected layer, whose weights are trained using simulation results, the actual classification accuracy will be significantly affected (92%). However, the Fourier kernel weights still bear the same representative information as simulation model, and that the fully connected layer weights need to be updated to fit the hardware convolution results, thus compensating for the quantitative discrepancies between the model used for training and hardware implementation. Therefore, we implemented an ulterior fine-tuning process, which uses the hardware convolution results to re-train the fully connected weights of the layer with a reduced number of training samples. In details, we perform the fine tuning which utilizes the existing knowledge learned by simulation model from full training set and learns a mapping from experimental results towards simulation results and compensate for it. (Section 8 Ref. 35) This approach proves to be particularly useful and the tuned hardware results accuracy shows a significant improvement (98%) compared with the one without fine-tuning (92%). Moreover, this fine-tuning approach which compensates from hardware-to-model discrepancies can be used if the optical engine is processing data in harsh environment conditions, for application such as super-resolution on object detection performance in satellite imagery, which can cause random misalignments.

| Model | MNIST | CIFAR |
|---|---|---|
| Space-domain convolution (full precision) | 98% | 63% |
| Simulation model (Fourier convolution) | 98% | 62% |
| Hardware model (without fine-tuning) | 92% | 25% |
| Hardware model (with fine-tuning) | 98% | 54% |

*Table I.* *Result of normal space domain convolution, our Fourier convolution simulation model, hardware model with and without fine-tuning. More details on Simulation results in Section 9 Ref. 35.*

For a more complex dataset, such as CIFAR-10, which comprises colour images of 10 classes, with 6,000 images per class, the inference accuracy for the simulated model is 62%, which is also close to the regularly used space-domain convolution model with full bit-precision, for similar neural network architecture (1 conv. layer) implemented in different technology, such as 1-layer electronic CNN or phase-only 4*f* schemes (accuracy of 63%). This is a promising result, since we show that in simulation our network with binarized kernel weights is able to obtain a (near) similar level of accuracy as normal space domain convolution using full precision features (32 bit). This can be explained by the effectiveness of the training of the 4*f* system, as well as the fact that there are more learnable parameters in the Fourier convolution, due to the larger kernel size compared with the space convolution version.

The ulterior degrees-of-freedom provided by the optical engine are considered to be 'free' since the convolution time in the optical system does not depend on the size of the kernel as long as the size is within DMD's resolution. After fine-tuning using a contained number (5000) of hardware results, the classification accuracy is a 54%, which is respectable given that it represents close to 90% from the nominal achievable results. (**Table I**)

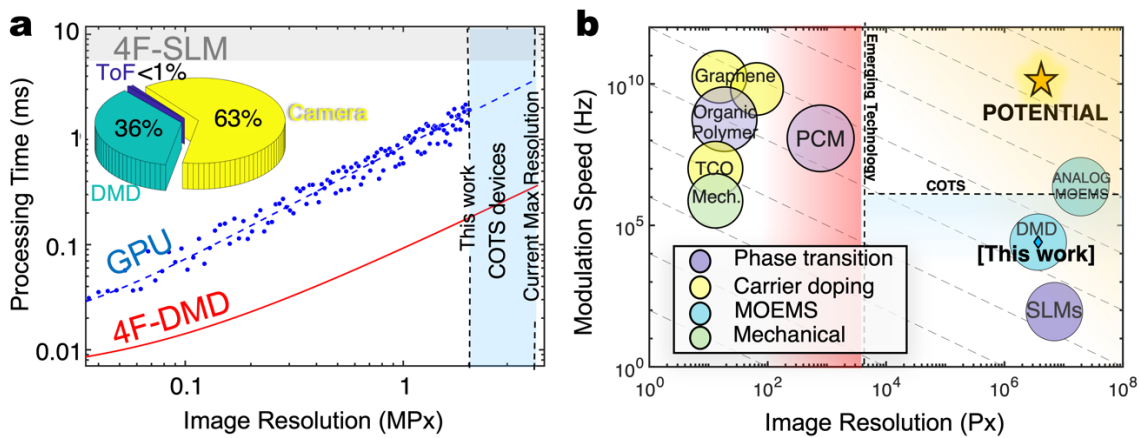

***Figure 3. Performance of the amplitude-only optical Fourier engine and its performance potential.*** ***a*** *Comparison of total processing time for performing a convolution as function of the image (matrix) resolution (expressed in Megapixels) comparing the Amplitude-only Fourier Filter (red solid line) to the P100 Nvidia GPU (blue-dashed line fitting, experimental data dots) and a 4f system based on Spatial Light Modulators (grey line). Here, we consider the convolution between two images (input and kernel) sharing the same pixel resolution expressed in MPx. The 2MPx mark set the current maximum resolution of the DMD of this experimental realization but does not represent a technological limit. Pie chart illustrates the breakdown of the latency for the DMD based 4f system when performing convolution. The overall latency consists of the DMD operation time (switching speed of the mirrors – green slice), camera integration time (yellow slice) and time of flight of the photon in the optical setup (violet slice).* ***b*** *Programmable electro-optic spatial light modulator grouped according to the functioning principle define processor performance defined by Matrix Size-Speed-Product (grey iso-performance lines). Exemplary, the 100x improvement over an SLM-based system (e.g. Optalysys) is a direct function of matrix-size and update rate: Carrier doping (Graphene[28,29], TCO[30]), Phase change (PCM[31], Organic Polymer[12], LCOS-SLM Gaea-2), MOEMS (Texas Instruments: 2MPx-DLP9000 and 4K-DLP660TE, Analog MOEMS[32]) and Electro-mechanical[33]; which can contemporary increase the throughput and lower latency of the proposed 4f system. The plot is tripartite into: Emerging technologies, COTS devices and potential hardware with GHz-fast, million-pixel electro-optic devices which can spatially modulate light for the next-generation information science and sensing.*

To provide some details regarding the efficiency and performance of this novel computing scheme based on 2Mpix DMDs, the AO-FF can perform convolutions between large matrices, in transform calculations, 10 times faster than a Nvidia P100 graphics card, commonly employed for high-performance computation, and more than 2 order of magnitude faster than architectures which exploits SLMs, while consuming similar power. In terms of efficiency (**Fig. 3a**), the largest portion of the energy consumption and processing time of our optical engine comes from the signal transduction step, from digital electronics to optical domain and vice versa. In our optical system, the processing time for performing an 8 bit convolution is given by the sum of all delays, including the generation of the patterns (DMDs), time-of-flight of the photons through the optical setup (ToF), detection by the CCD camera (Camera), and ultimately being transmitted for subsequent software processing. For a 2 Megapixel (MPx) 8-bit input and kernel images, the largest contribution latency is given by the camera acquisition time, followed by the DMD switching speed. The propagation time is negligible since, considering the 4f distances into play and the optical tools, it amounts for few nanoseconds. Whereas, the acquisition time of the high-speed camera is function of the resolution of the image to be detected and represents the bottleneck of this current implementation. Higher speed camera can ameliorate the processing time by a factor of 2, keeping the same DMDs speed and resolution.

Looking at the future potential of this 4f-based hybrid accelerator paradigm, developments of faster and higher-resolution spatial modulators and high-speed detection mechanisms are crucial to the advancement towards the implementation of intelligent functionalities (**Fig. 3b**). For instance, higher resolution DMDs (4k resolution) and cameras would lead to an even increased parallelism (16 times the current throughput) compared to our prototype. Interestingly, at the research level, analogue version of MOEMs can reach high modulation speed (~10MHz) without trading off pixel resolution (~10 MPx)[32]. Using the analogue MOEMS, for spatially modulating the optical signal, in combination with the ultra-high-speed camera (MHz, >4K resolution), for converting the filtered signal in the electric domain, could improve the throughput of the system by about 4 orders of magnitude. However, for this configuration, the electronic interface will be the bottleneck of the system, which has to be able to deliver the patterns and acquiring data with an overall bandwidth of tens of ~100 Tera operations per second. Nonetheless, our AO 4f optical processor demonstration paves the way to future realizations; for instance exploiting emerging technology components such as micrometer-thin metalenses, GHz fast reprogrammable metasurfaces, and high-speed photodiode arrays would yield to highly competitive footprint, while augmenting the computation throughput up to exa-operations per seconds, without trading off in terms of power consumption. However, at the current stage, these components are still challenged in terms of matrix resolution and achievable phase contrast[11,12]. These devices necessitate materials and device configurations which can provide efficient light matter interactions, CMOS compatibility, straightforward and stark tunability, and sufficient maturity to be scaled up.

## Conclusions

In summary, we have demonstrated an amplitude-only electro-optic Fourier filter engine with high speed programmability and throughput. The dynamic Fourier filtering is realized using digital micromirror devices, both in the object and Fourier plane of an optical 4f system. As a proof-of-principle demonstration, we constructed a Neural Network which uses, as convolutional layer, the electro-optical convolutional engine for classifying handwritten digits (MNIST) and colour images (CIFAR-10). We trained the network off-chip, using a detailed physical model which describes the electro-optical system and its nonidealities, such as optical aberrations and misalignments. After experimentally validating the model and retraining the following fully-connected layer to compensate for values discrepancies, we obtained a classification accuracy of 98% and 54% for MNIST and CIFAR-10, respectively, with a throughput up to 1,000 convolutions per seconds between two 2MP images, which is one order of magnitude faster than the state-of-the-art GPU. Additionally, our scientific contribution emphasizes that the information loss and inaccuracies deriving from neglecting the phase of the optical wave front can be compensated-for by the degree of robustness provided by neural network training, which yields intelligent classification, at high accuracy as the one obtained by phase-only optical engine, while featuring 2 orders of magnitude faster programmability. The system can also be used to filter images of smaller resolution in parallel, and by exploiting ad-hoc electronic I/O interface, emulate deeper neural networks reaching high number of connections and millions of neurons. This paradigm and hardware implementation of the optical engines for artificial neural networks is a promising alternative to other machine learning architecture since they can avail parallel computing capability and power efficiency inherent to optical systems. Our results, reported for different inference tasks, indicate the potential that our intelligent information processing scheme could open new perspectives of a flexible and compact platforms which could be transformative for diverse applications, ranging from image analysis to image classification and super-resolution imaging on unmanned aerial vehicle , and may also enable high bandwidth free-space communication in data centres, intelligently pre-processing data locally at the edge of the network.

**Acknowledgments:** We thank Prof. Aydin Babakhani, Prof. Seth Bank, Dr. Hamed Dalir, Prof. Tarek El-Ghazawi, Prof. David Pan, and Prof. Chee Wei Wong of the "Photonic Convolutional Processor for Network Edge Computing" team for the insightful discussions.

**Funding:** This work was supported by Office of Naval Research Electronic Warfare Program under award number N00014-19-1-2595. V.S. is supported by the Advanced Computing Program (ACI) under the Army Research Award number W911NF1910468.

**Author contributions:** V.J.S. and M.M. envisioned the idea of an amplitude only Fourier Convolutional engine for deep learning, V.J.S. and P.G. acquired the funds, and supervised the project. M.M. developed the relevant theories and analyses for the project. M.M. and Z.H. designed the experimental setup and conducted the free-space experiments. S.L. designed and trained the amplitude only Fourier neural network and performed the relevant tests and benchmarking. M.M., P.G., S.L. P.H., J.G. and V.J.S. discussed the results and contributed to the understanding, analysis, and interpretation of the results. MM wrote the first draft of the manuscript, and P.G., S.L. Z.H., J.G., R.C., P.M.B. and V.J.S contributed to writing subsequent drafts of the manuscript.

**Competing interests:** The authors declare no competing interests.

**Data and materials availability:** All data needed to evaluate the conclusions in the paper are present in the paper or the supplementary materials.

# Supplementary Online Information for

# Massively Parallel Amplitude-Only Fourier Neural Network

*Mario Miscuglio[1], Zibo Hu[1], Shurui Li[2], Jonathan George[1], Roberto Capanna[3], Philippe M. Bardet[3], Puneet Gupta[2], and Volker J. Sorger[1].\**
\* *sorger@gwu.edu*

[1]*Deptartment of Electrical and Computer Engineering, George Washington University, Washington DC, DC, USA*
[2]*Deptartment of Electrical and Computer Engineering, University of California, Los Angeles, CA, USA*
[3]*Department of Mechanical and Aerospace Engineering, The George Washington University, Washington DC, USA*

## Outline: Section (S)
**S1: System Data I/O**
**S2: Applications**
**S3: Physical Modeling**
**S4: Structural Similarity Experimental Verification**
**S5: Parallelization Strategies**
**S6: CNN Training**
**S7: Network Structure**
**S8: Fine tuning for Non-idealities**
**S9: Simulation Results**

**Section 1 – FPGA Controlled unified I/O Interface for emulation of a convolutional neural network**

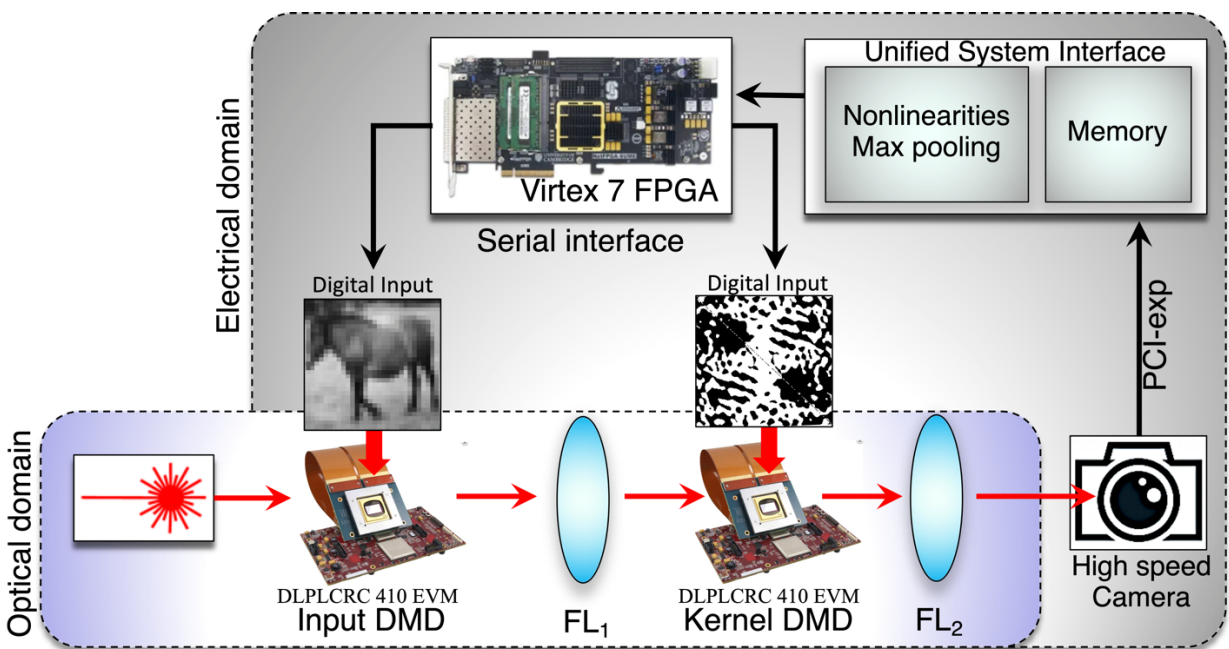

***Figure S1. System architecture for performing convolution in streaming.*** *A PC interfaces (through a PCIe) an FPGA (Virtex 7) which controls through a serial interface the DLP (DLPLCRC 410 EVM, Texas Instrument) boards. Notice that the FPGA can also be used for generating the pattern directly. The output of the camera (IDT Y7-S3) is connected to the PC using an ethernet cable. The output of the convolutional layer can be stored or postprocessed electronically by the PC. The FPGAs sends the trigger signal for synchronizing the refresh rate of the DMDs and*

*acquisition time of the high-speed camera. The high-speed camera can be set to record a predetermined time window at a given framerate. In this way the same optical hardware can be used for successive convolutional layer.*

## Section 2 – Applications

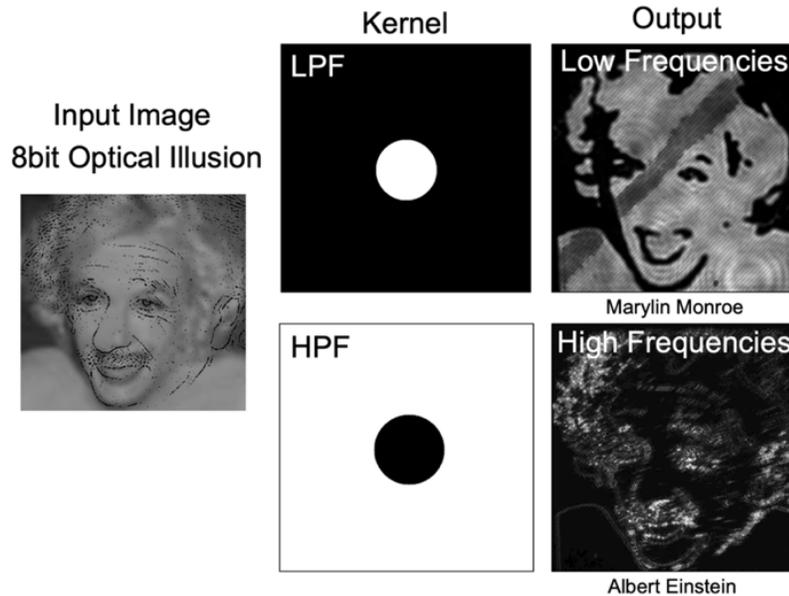

***Figure S5. Detection of optical illusions and augmented visual perception on 8bit images.*** *The higher switching speed of the DMDs compared to SLMs allows for fast detection of information camouflaged in images. The engine can be used for augmented perception processing filtering images at high speed. Here, as a demonstration of this potential, we experimentally demonstrate the augmented visual perception by filtering a famous optical illusion. The illusion consists of two 8bit portraits overlapped. In details, Marilyn Monroe is in low spatial frequency component and Albert Einstein is in high frequency component. A spatial filter which can be updated at 20kHz can detect the 2 faces instantaneously.*

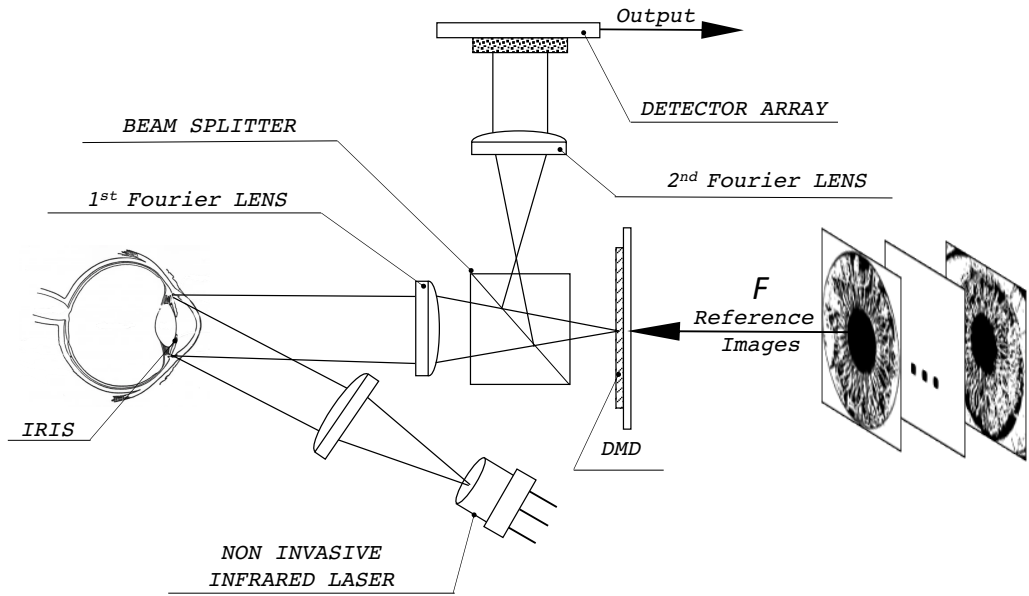

***Figure S6. High speed retina recognition using DMD-based 4F system.*** *A noninvasive low power infrared laser, after being collimated, illuminates the eye of the user. The reflected light is transformed into its spatial frequency component passing through a Fourier lens (1st Fourier Lens) and 2 focal distances. At the Fourier plane, FT image interacts with the pattern generated by a DMD, loaded with different signatures updating at a speed of 20kHz. The*

*output is inverse transformed (2ⁿᵈ Fourier Lens) and image by the camera for subsequent post processing (fed to Fully Connected Neural Network)*

**Section 3 – Physical Modelling**

The first step is modelling the DMD magnitude transfer function (MTF) which represents the capability of transferring the modulation depth of the input to the output signals at a specific spatial frequency, which represents the degree of fidelity of the digital signal (**Fig. S3a**). Next we will proceed describing the physical model of the coherent amplitude only Fourier filter. When describing the system, we will refer to the coordinate system shown in **Fig. S3b**, in which we perform segmentation on the photo of the Science and Engineering Hall (GWU). For the DMD, each pixel is considered as tiny square mirror with a hole etched at the centre which does not reflect light (pin on which the micromirror is hinged), and each mirror is separated from the neighbouring mirrors by a small distance. When a driving electrical signal is applied to the DMD, an electro-static force is created between the electrodes of the selected mirror, so that it is tilted to deliver the illuminating light into the optical system (22.5º). OFF mirror will reflect the illuminating light out of the optical system. The DMD module used in the system is constituted of an array of up to 1920x1080 micromirrors with full addressing circuitry. To address the pixel structure, each mirror can be individually driven to rotate by applying a potential difference between the mirror and the addressing electrode. The response time of each mirror is 10 μs, and the addressing voltage is 5 V. The pixel pitch of the micromirror array of the DMD is about 17 μm. Each pixel is a square micromirror of dimensions 16μm×16μm, and with an etched hole of 1μm diameter at the centre. Therefore, the fill factor *r* is approximately equal to 16/17, and the normalized radius of the hole, $r_c$, is 0.5/17. For this reason, the algorithm performs a 17x17 pixel expansion of the input image associating the pattern of the mirror for modelling the optical image fidelity. The MTF takes into the account also imperfect contrast of the ON-OFF ratio which can be altered for each pixel in an 8bit resolution depth. (**Fig. S3 c-i**) The model follows in characterizing the non-uniform phase induced by the orientation and tilted angle of the micromirror array with respect to the propagating beam direction. (**Fig. S3 c-ii**). The electric field which accounts for the tilting angle of the DMD and its orientation in space is obtained by elementwise multiplication of the field patterned by the 1ˢᵗ DMD and phase term proportional to the distance from the centre and tilting angle $\theta$.

$$E_{Ph\_correction} = E_{DMD}\boldsymbol{\phi}$$   Eq. 1

, where the element of the matrix $\phi_{i,j}$ is the phase term, computed $\phi_{i,j} = e^{-i(\sin\theta\, d_{i,j}2\pi/\lambda)}$, where $d_{i,j}$ is the physical distance between the centre of the DMD, considering the 45º in-plane rotation.

The wavefront of the input image g($x, y$) passing through the lens is Fourier transformed at a distance 2*f* from where it was originated. $G(u, v)$ according to Fresnel Integral:

$$G(u, v) = \iint\limits_{-\infty}^{+\infty} g(x, y)\exp[-i2\pi(ux + vy)]dx\,dy$$   Eq. 2

The change of coordinates from the spatial frequency domain to the real space in the Fourier plane is function of the wavelength and focal length $(u, v) = \left(\frac{x''}{\lambda f}, \frac{y''}{\lambda f}\right)$. Next, we take into account the lens aperture, and aberrations to the wave-front according to characteristic Seidel coefficients. The seidel coefficients considers potential Defocus, Spherical, Coma, Astigmatism and Field curvature Distortion of the lens and modify the phase term (**Fig. S3 c-iii**), as shown in Eq.3.

$$
\begin{aligned}
W(\hat{u}_0;\hat{x},\hat{y}) = &\, W_d\left(\hat{x}^2 + \hat{y}^2\right) + W_{040}\left(\hat{x}^2 + \hat{y}^2\right)^2 \\
&+ W_{131}\hat{u}_0\left(\hat{x}^2 + \hat{y}^2\right)\hat{x} + W_{222}\,\hat{u}_0{}^2\,\hat{x}^2 \\
&+ W_{220}\,\hat{u}_0{}^2\left(\hat{x}^2 + \hat{y}^2\right) + W_{311}\,\hat{u}_0{}^3\,\hat{x}.
\end{aligned}
\qquad \text{Eq. 3}
$$

$$
H = A(u,v)\,\mathrm{e}^{-\mathrm{i}k\mathbf{W}}, \qquad \text{Eq. 4}
$$

being $u_0$ normalized image height, defined along the $u$ axis in the imaging plane, A is the circular function which defines the circular aperture, given in terms of exit pupil size and pupil distance. The aberrated wave-front of the Fourier transform is obtained by multiplying it with the H functions. In this view, for a fixed wavelength, the lens is selected with respect to its focal length which dimension the Fourier transform. Our goal here is to exploit the entire resolution of the 2$^{nd}$ DMD for having the max degree of freedom in selecting and filtering the spatial frequency of the input images without losing frequency components.

$$
G'(u,v) = H(u,v) \cdot G(u,v) \qquad \text{Eq. 5}
$$

The interaction with the second DMD is obtained by performing a pixel-wise multiplication between the 2$^{nd}$ DMD pattern and the impinging wave front, according to Huygens' Principle.

$$
G''(u,v) = G_{DMD_2}(u,v) \cdot G'(u,v) \qquad \text{Eq. 5}
$$

The resulting beam is inverse Fourier transformed obtaining the convolution in the real space (with flipped axis). This step considers the aberration and *f number* of the 2$^{nd}$ Fourier lens with the rationale of having an image in the image plane of the same size of the CCD sensor (**Fig. S1 c-iv**).

$$
g_{out}(x',y') = \iint\limits_{-\infty}^{+\infty} G''(u,v)\exp[-i2\pi(ux' + vy')]du\,dv \qquad \text{Eq. 2}
$$

In the algorithm, the CCD camera accomplishes the dimensionality reduction integrating the optical power (square optical intensity) mapping each expanded 17x17 super-pixel to a single pixel. (**Fig. S1 c-v**). It is worth mentioning that the algorithm used for modelling the system can be used for similar 4$f$ system which uses miniaturized reprogrammable metasurfaces and flat diffractive metalenses. In that case, the characterization of the optical tools and their inaccuracies would have even a greater impact to the results provided by the optical engine.

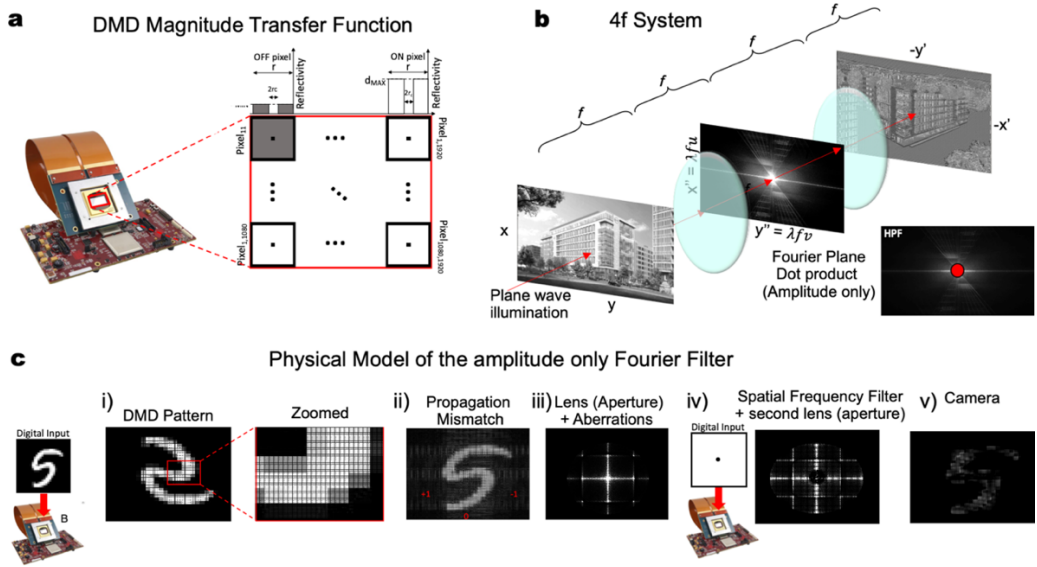

*Figure S3. Physical modelling of the amplitude only convolutional engine. **a.** DMD magnitude transfer function considers the pixel dimension (mirror size and pitch), resolution, on-off contrast and the effective pixel dimension (fill-factor) **b.** 4f system performing image filtering (Segmentation) of an 8bit resolution (greyscale) 2MPx Image. A digital micro mirror device modulates the expanded beam coming from a HeNe laser as a GW Science and Engineering Hall building. The beam is then transformed in the Fourier plane from the first lens and convoluted with a spatial filter, which cancels out the low frequency components of the image. After being inverse transformed in the real space from the second lens the real convoluted image is acquired by a high-speed camera. The image acquired results into the outline of the building. **c.** Physical model description of the Amplitude Only (AO) Fourier Filter considers that i) The input pattern is generated considering the DMD Magnitude Transfer Function (MTF), therefore including contrast (reflectivity) and effective pixel size; ii) the slanted angle at which the DMD is oriented, by assigning a non-uniform phase delay for the pixel (function of the plane inclination). iii) the lens aperture and aberration using Seidel Polynomial. iv) spatial Frequency filter (pixel wise product with the 2nd DMD pattern) and second lens aperture and aberration (and opposite non-uniform phase given by the orientation and slanted angle of the 2nd DMD) v) Camera squaring the intensity of the electric field and eventually reduce the pixel size (if different pixel resolution than DMD).*

## Section 4 – Structural Similarity Experimental Verification

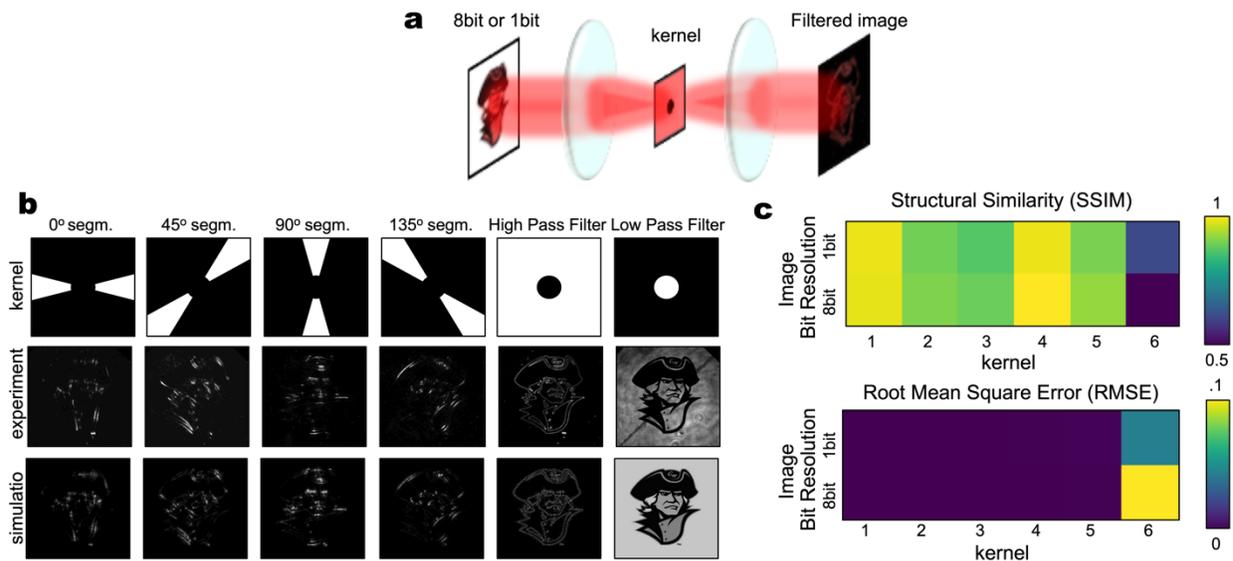

*Figure S4. Performance and experimental validation of the of the amplitude only Fourier filter. a Segmentation of the GW mascot logo (the Colonial) b Several spatial frequency filtering of the 8bit (grey scale) high-resolution image of the GW mascot logo and comparison with simulations. c Comparison between numerical simulation and experimental results for several kernels. Goodness of the agreement between experiment and model is expressed in terms of Structural Similarity (SSIM) and Root mean square error (RMSE) of the images in c.*

## Section 5 – Parallelization Strategies

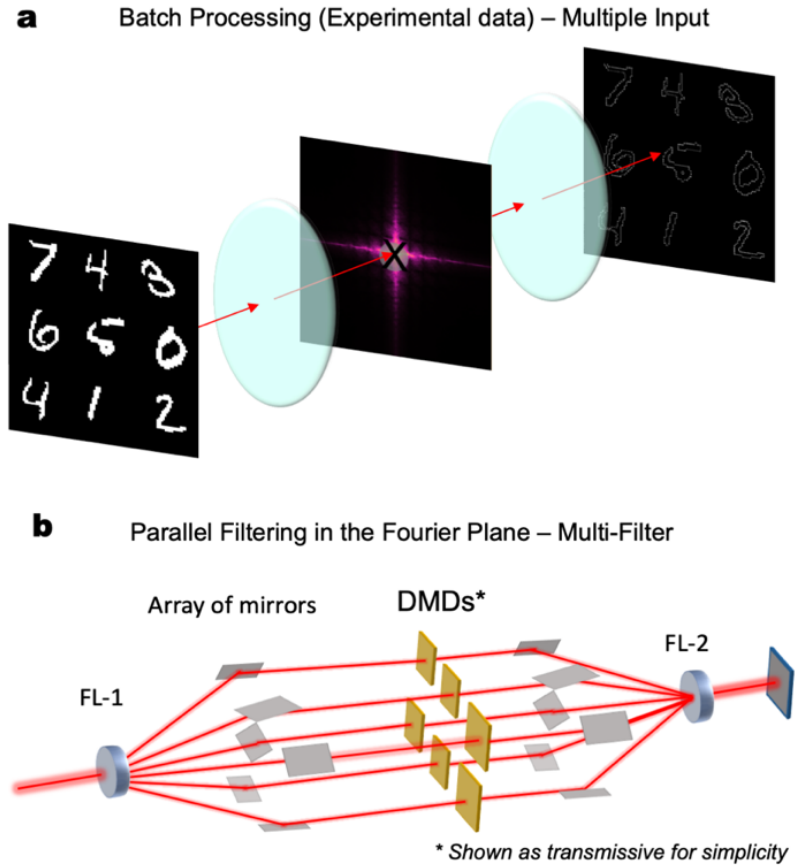

*Figure S5. Parallelization techniques. a. Experimental implementation which exploits the large resolution of the DMD for processing in parallel multiple 28×28 input images tiled together in the object plane. The frequency components are processes in parallel using a spatial filter in the Fourier plane. The result of the filtering is imaged after being inverse Fourier transformed into the real space by the second Fourier lens and imaged by the camera. b. Conceptualization of a multiple frequency filters applied into the Fourier plane to the same input image. The Fourier transformed image is directed to different non overlapping portion of the DMD (or different DMDs) using opportune beam splitters and array of mirrors (using also the higher diffraction orders). In the Fourier plane, the Fourier transforms of the input image will impinge on the different DMDs obtaining the pixelwise product. Ultimately each product is inverse Fourier transformed (using a lenslet array) and imaged by different sensors. Alternatively, the product can be integrated by the same sensor (dimensionality reduction).*

## Section 6 – CNN Training

<u>Datasets:</u> We used two classic datasets for image recognition to demonstrate the learning capability of this system as well as to evaluate its performance. MNIST dataset is a widely used dataset and is about recognizing hand-written digits. We choose MNIST dataset as a proof of the system's learning capability.

CIFAR-10 is another popular image classification dataset and it tests the network's ability to classify objects includes different types of animals and vehicles. It posts a more challenging problem than MNIST and it is used for evaluating and comparing the model's performance. The CIFAR-10 dataset uses RGB images which consists three colour channels whereas the MNIST dataset uses grayscale images. For CIFAR-10 input channels are processed individually and their convolution results are summed up.

_Training setup_**:** We trained our model using PyTorch platform and implemented custom Fourier convolution layer in order to simulate the physical system. We also created some custom functions for complex number operations, as PyTorch does not support complex datatype natively. All functions are written using PyTorch's built-in functions so that Pytorch's autograd function can be directly applied for backpropagation during training.

_Fourier domain training:_ We adopted the concept of FFT-based Fourier domain training [25],together with the refined hardware model to accurately simulate the complete process and learn the kernel weights. The kernel values, which are the learnable parameters of the convolutional layer, are initialized directly in the Fourier domain. By doing so the kernels don't need to be transformed into Fourier domain like in [26,27], which fits our physical model better. We did not constrain our model to simulate what traditional space-domain convolution does exactly, which is convolving the input with real-valued filters and equivalent to complex kernels in Fourier domain. Since the DMD restricts the kernel values to be real and binary, if the kernels are trained in space domain and further transformed in to Fourier domain, their absolute values need to be taken before the binarization which will leads to more information loss and lower performance. Hence, we initialized the kernel weights in the Fourier domain so that they match directly to the DMD cell values, therefore the learned weights can be loaded into kernel DMD without further modifications.

_Binarization_**:** For DMD using 1-bit mode, its cell values are either zero or one, hence during training stage the kernel weights need to be binarized in order to compensate the restriction on cell values. The binarization scheme we used here is slightly different than the common sign function approach mentioned in [34].We modified the sign function scheme slightly, all positive weights are replaced with 1 while negative weights are replaced with 0, to fit the requirement of DMD. The mathematical formula is:

$$x_b = \begin{cases} +1 \ if \ x > 0, \\ 0 \ \ if \ x \leq 0, \end{cases}$$

Where $x_b$ is the binarized weight and $x$ is the original real-valued weight. This binarization scheme is deterministic and simple to implement. The binarization process only applies during forward pass and in backpropagation this step is skipped. In our system, only the kernel weights of the convolution layer are binarized, the activations and weights for fully connected layers remain to 32 bits, so as the outputs of convolution.

_Pre-processing_**:** In order to let the trained network, fit the DMD system properly, some modifications and pre-processing need to be implemented during training process. Input images are zero-padded from original size to $208 \times 208$ before the Fourier transform and the kernel size is also set to $208 \times 208$. The reason behind the padding is that during actual inference the input needs to be expanded by four times (each pixel converts to a 4x4 block with same value) so that the maximum spatial frequency component can be captured by the kernel DMD (considering the change of coordinates in the Fourier plane). The effective DMD resolution is 832x832, therefore the original size before expansion should be zero-padded to $208 \times 208$.

## Section 7 – Network Structure

_Network structure:_ The simulation model is a Fourier convolution network with a single Fourier convolution layer and one fully connected layer. Max-pooling layer is applied after the convolution layer for down-sampling and ReLU is used as the non-linear activation function for fully connected layer. For convolution layer, since the camera (which captures the light intensity) is already a source of non-linear function, no extra activation function is used for convolution layer. A batch normalization layer is applied after the convolution layer for better training results. Since the inputs are padded to 208 × 208, the kernel size is also 208 × 208 as their size after Fourier transform need to be matched in order to perform the point-wise multiplication in Fourier domain. The Fourier convolution layer consists of 16 binary frequency-domain kernels which generate 16 corresponding output feature maps of size 208 × 208. Since the original input size before padding is 28 × 28(MNIST)/32 × 32(CIFAR), the actual convolution result should be the centre 28 × 28/32 × 32 of the output feature map generated by the Fourier convolution layer, the other region of the 208 × 208 output are basically zeros. Thus, to reduce the network size and alleviate memory usage, only the centre region of the output feature maps are extracted and processed, before feeding into fully connected layer. **Figure S7** shows the overall network structure of our Fourier convolution model.

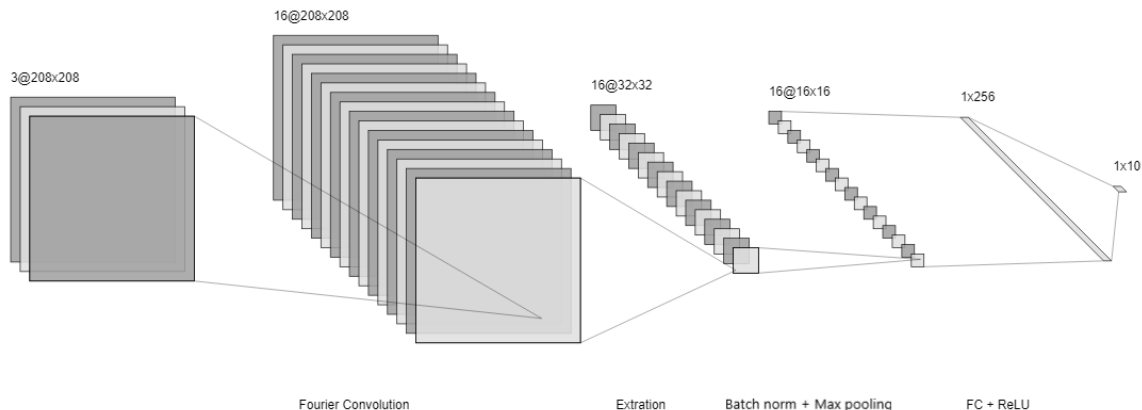

*Figure S7: Network structure for CIFAR-10 dataset.*

**Section 7 – Fine Tuning to Compensate Non-idealities**

Since our simulation model already taken into account some non-ideality of the hardware system, the hardware convolution results match the simulation result quite well qualitatively, their shapes are almost identical. Still, the match is not perfect quantitatively, due to several factors including alignment and non-ideal camera. The exact pixel values of hardware results differ from the simulation results, thus if the hardware convolution results are feed into fully connected layer whose weights are trained using simulation results, the actual classification accuracy will suffer. However, since the shape of hardware results match with the simulation result, it means that the Fourier kernel weights learned using simulation model is valid, the hardware convolution result still has same representative power as simulation model, just the fully connected layer weights needs to be updated to fit the hardware convolution results. Therefore, we implemented a fine-tuning process which uses the hardware convolution results to re-train the fully connected layer's weights. This approach proves to be useful and the tuned hardware result's accuracy shows a significant improvement compared with the one without fine-tuning.

The fine-tuning approach uses the first fully connected layer (FC1) to perform a mapping from experiment results to simulation results. The goal is to minimize the difference between outputs of simulation and experimental data's FC1, so that the weights trained in simulation of the output layer (FC2) can be used directly. Least square (L2) error between simulation and experimental FC1 outputs before ReLU is used as the loss function for fine-tuning.

Therefore, the FC1 layer is used to learn the non-linear mapping and produce outputs as close as possible to FC1 outputs of the simulation. Compared to completely retrain the two FC layer using cross entropy loss, this approach utilizes the existing knowledge learned by simulation model from full training set and learns a mapping/correction towards it; which makes this approach performs better when the number of experiment data available for fine-tuning is limited.

**Section 8 – Simulation Results (Table S1)**

_Results:_ For MNIST dataset, our model achieves 98% accuracy in simulation, which is same as a single layer space-domain convolution neural network with 16 filters using full precision. The result indicates that the proposed system is able to function as a normal convolution neural network. For CIFAR-10, the simulation accuracy is 62%, which is also very close to the normal space domain model with full precision using similar structure (accuracy of 63%). This is promising result since we showed that in simulation our network with binarized kernel weights is able to get same level of accuracy as normal space domain convolution using full precision. The close performance between binary Fourier kernel and full precision space domain kernel might be explained by the effectiveness of Fourier training, as well as the fact that there are more learnable parameters in the Fourier convolution version. The extra learnable parameters in Fourier convolution model is due to the larger kernel size compared to space convolution version and they are considered to be 'free' since for optical system the convolution time does not depend on with size as long as the size is within DMD's resolution.

The hardware results without fine-tuning drops to 25% compared to regular space-domain convolution and simulation results. This is due to the fact that CIFAR is a harder dataset compared with MNIST and the patterns are significantly more complex, which can be affected more considerably from alignment and camera errors. After fine-tuning using a contained number (5000) of hardware results, the classification accuracy is further refined up to 54%, which is much closer (90%) to the nominal simulation results.

_Training detail_: We use Adam as the optimizer and set the learning rate to default value (0.001). For MNIST the model is trained for 10 epochs and for CIFAR-10 it is trained for 15 epochs.

_Accuracy comparison_: In order to evaluate the model's performance, we compare the result of simulation model to both the space domain convolution model and ideal Fourier convolution model (implemented directly using FFT and IFFT, did not model the hardware imperfections). For all models the number of filters is set to 16 and the $208 \times 208$ input padding is not implemented for benchmark purposes. The result shows that the simulation model achieves same performance as ideal model for both benchmarks. The Fourier convolution model using full precision performs slightly better than space convolution model, most likely due to the relatively large kernel size expands the parameter space, so the network is able to learn better.

The ideal model performs Fourier convolution, but does not take into account the actual hardware model while the simulation model simulates the actual $4f$ system. Note for both table the results are for Fourier kernel size $32 \times 32$, which is different from the final adopted model.

The performance loss of binarized CIFAR-10 versions is mainly due to the binarization of kernel weights and the fact that CIFAR-10 is a harder benchmark than MNIST and using single layer with binarized weights sacrifices accuracy. However, we showed that if we switch the DMD from 1-bit mode to 2-bit mode, only the addition of a single bit in kernel weights would result in a boost of accuracy from 55% to 61% in simulation. This result is promising since as it shows that with small throughput loss, we can get similar performance as the space-domain counterparts with full precision.

**Table S1.** _Inference Results based on physical model of the optical system and electrical max-pooling, flattening, and FC-layers._

| Model | MNIST | CIFAR-10 |
|---|---|---|
| Space convolution, full precision | 98% | 63% |
| Fourier convolution, ideal model, full precision | 98% | 64% |
| Fourier convolution, ideal model, 2-bit kernel | 98% | 61% |
| Fourier convolution, ideal model, 1-bit kernel | 98% | 54% |
| Fourier convolution, simulation model, full precision | 98% | 65% |
| Fourier convolution, simulation model, 2-bit kernel | 98% | 61% |
| Fourier convolution, simulation model, 1-bit kernel | 98% | 54% |

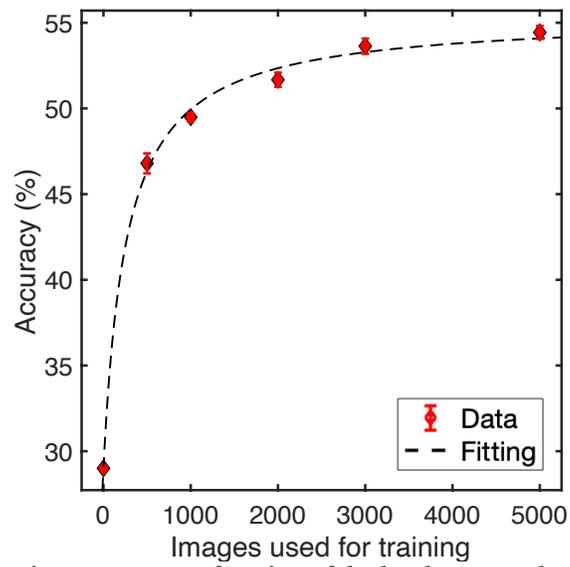

*Figure S8. Inference accuracy improvement as function of the hardware results used to finely training the FCNN.*

*Video 1 – Output of the experimental implementation of the convolutional layer when performing classification on MNIST dataset at 20 kHz. youtube.com/watch?v=s7L7yFlx44A*